\documentclass[twocolumn,showpacs,preprintnumbers,amsmath,amssymb]{revtex4}

\usepackage{graphicx}
\usepackage{dcolumn}
\usepackage{bm}

\newcommand{\Piz}{\ensuremath{\pi^\circ}}
\newcommand{\gprim}{\ensuremath{\gamma^\prime}}
\newcommand{\pigp}{\ensuremath{\Piz{} \gamma^\prime p}}
\newcommand{\Rpigp}{\ensuremath{\gamma
p \rightarrow \Piz{} \gamma^\prime p}}

\newcommand{\Rpip}{\ensuremath{\gamma
p \rightarrow \Piz{} p}}

\newcommand{\GDDg}{\ensuremath{\Delta \rightarrow \Delta \gamma^\prime}}
\newcommand{\muD}{\ensuremath{\mu_{\Delta^+}}}

\begin{document}


\boldmath
\title{The reaction \Rpigp{} and the 
magnetic dipole moment of the $\Delta^+(1232)$ resonance}
\unboldmath

\author{
  M. Kotulla$^1$,
  J.~Ahrens$^2$,
  J.R.M.~Annand$^3$,
  R.~Beck$^2$,
  G.~Caselotti$^2$
  L.S.~Fog$^3$,
  D.~Hornidge$^2$,
  S.~Janssen$^1$,
  B.~Krusche$^{4}$,
  J.C.~McGeorge$^3$,   
  I.J.D.~McGregor$^3$,
  K.~Mengel$^1$,
  J.G.~Messchendorp$^1$,
  V.~Metag$^1$,
  R.~Novotny$^1$,
  M.~Pfeiffer$^1$,
  M.~Rost$^2$,
  S.~Sack$^1$,
  R.~Sanderson$^3$,
  S.~Schadmand$^1$,
  D.P.~Watts$^3$
}
\affiliation{%
  $^1$II. Physikalisches Institut, Universit\"at Gie{\ss}en,
      D--35392 Gie{\ss}en, Germany\\
  $^2$Institut f\"ur Kernphysik, Johannes-Gutenberg-Universit\"at Mainz,
      D--55099 Mainz, Germany \\
  $^3$Department of Physics and Astronomy, University of Glasgow,
      Glasgow G128QQ, UK \\
  $^4$Department of Physics and Astronomy,
      University of Basel, CH-4056 Basel (Switzerland)
}%

\date{\today}

\begin{abstract}
The reaction \Rpigp{} has been
measured with the TAPS calorimeter
at the Mainz Microtron accelerator facility MAMI for energies between
$\sqrt{s}$ = 1221--1331 MeV.
Cross sections differential in angle and energy have been determined 
for all particles in the
final state in
three bins of the excitation energy.
This reaction channel provides access to the magnetic dipole moment
of the $\Delta^{+}(1232)$ resonance and, for the first time, a
value of 
$\mu_{\Delta^+} = (2.7_{-1.3}^{+1.0}(stat.) \pm 1.5 (syst.) \pm
3(theo.))\;
 \mu_N$
has been extracted.
\end{abstract}

\boldmath
\maketitle
\unboldmath

The complex structure of the nucleon is reflected in its rich excitation
spectrum. Attempts to unravel the baryon structure have led to an
impressive determination of 
the properties of the nucleon, e.g. polarizabilities,
magnetic moments, and more general form factors. 
Additional and substantial insight  
in the parton structure of the nucleon has been gained
through deep inelastic electron scattering.
In contrast to that, the
knowledge of the nucleon's excited states is limited to the mass of the lowest
resonances and its (iso)spin quantum numbers. However, 
to test the modelling of internal degrees of freedom of the excited
states, measurements of static properties are required. 
In particular, the properties of the 
$\Delta(1232)$ resonance are of
considerable interest because of its prominent position in the excitation
spectrum.

In this context, the magnetic moment is
an important observable for testing theoretical baryon
structure calculations. Different predictions for the magnetic moment were
made in several calculations \cite{leinweber:muD,butler:muD,kim:muD,aliev:muD}.
The magnetic moments of the octet of baryons ($N,\Lambda,\Sigma,\Xi$) 
of the SU(3) flavor symmetry classification
are known very accurately through spin precession
measurements. However, for the decuplet baryons, only the $\Omega^-$ magnetic
moment has been
determined as the lifetime of the other decuplet members is
too short for this technique. If SU(3) flavour symmetry were to hold, the
$\Delta$ and the nucleon would be degenerate in mass 
and their magnetic moments
related through $\mu_\Delta = Q_\Delta \mu_p$, where $Q_\Delta$ is the $\Delta$
charge and $\mu_p$ the proton magnetic moment. However,
structure calculations predict significant deviations from this
SU(3) value \cite{leinweber:muD,butler:muD,kim:muD,aliev:muD}.

\begin{figure}
  \includegraphics[width=0.56\columnwidth]{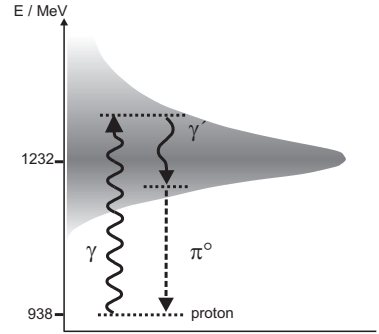}
  \caption{Method to study the static electromagnetic
	properties of the $\Delta^{+}(1232)$ isobar. The $\gamma^\prime$
        transition carries the information of the magnetic moment of
  the $\Delta^+$.}
\label{fig:m1gamma}
\end{figure}

It has been proposed that the electromagnetic structure
of the $\Delta$ can be determined 
by measuring a $\gamma$-transition within the resonance
\cite{kontra:delta}.
This method is depicted in Fig.~\ref{fig:m1gamma}, which shows
an energy level diagram with the proton (nucleon) as the ground state and
the $\Delta$ as the first excited state.
The $\Delta$ structure can be probed by exciting the proton
to a $\Delta$, which then emits a real photon and subsequently
decays into a nucleon and a pion.
Spin and parity conservation require that the lowest order
electromagnetic transition is magnetic dipole (M1) radiation.
This \GDDg{} amplitude is proportional to
$\mu_{\Delta^+}$ and was recently investigated in theoretical calculations
\cite{macha:pi0gp,macha:pi0gp_err,drechsel:pi0gp1}.
The next allowed multipole is the electric
quadrupole (E2) transition, but this amplitude vanishes in the
limit of zero photon energy because of time reversal symmetry 
\cite{drechsel:pi0gp}. The E2/M1 ratio of the transition
amplitude $N \rightarrow \Delta$  has been measured to be very small, 
approx 0.025 \cite{beck:rem}, which
leads to the assumption that the quadrupole deformation of the $\Delta$ is
very small. The 
magnetic octupole (M3) transition is
suppressed by two additional powers of photon momentum.
Hence, the measurement of the reaction
\Rpigp{} provides access to
\muD{}. Unfortunately this final state
can also result from bremsstrahlung radiation of the
intermediate $\Delta$ and the proton.
These
contributions are of the same order as the \GDDg{} transition of
interest.
Nonresonant contributions are expected to play a minor
role, since the partial wave decomposition of the related elastic 
channel \Rpip{} shows 
the dominance of the $\Delta$ resonant reaction process \cite{maid:pi0}.
The reaction channel $\gamma p\rightarrow\pi^+\gamma^\prime n$ is in
that sense less favorable for extracting 
the magnetic moment of the $\Delta^+$ isobar.
An accurate 
theoretical description
of all processes  contributing to \Rpigp{} 
is crucial for extracting a precise value for $\mu_{\Delta^+}$.

The magnetic moment of the ${\Delta^{++}}$ isobar was extracted
in a similar
way from the reaction $\pi^+ p\rightarrow\pi^+\gamma^\prime p$.
Two experiments at the University of California (UCLA) \cite{nefkens:muD}
and the Schweizerisches Institut f\"ur Nuklearforschung (SIN,
now called PSI) \cite{bosshard:muD}
have been performed and as a result of
many theoretical analyses of these data the
Particle Data Group \cite{pdg_a:2000} quotes
a range of $\mu_{\Delta^{++}}$ = 3.7--7.5~$\mu_N$ (where $\mu_N$ is
the nuclear magneton). The large uncertainty in the extraction of 
$\mu_{\Delta^{++}}$ is due to the strong contribution of $\pi^+$
bremsstrahlung and model dependencies. In the reaction channel described
here, the bremsstrahlung contributions are much weaker.

The reaction \Rpigp{} was
measured at the electron accelerator Mainz Microtron
(MAMI) \cite{walcher:mami,ahrens:mami} using the
Glasgow tagged photon facility \cite{anthony:tagger,hall:tagger}
and the photon spectrometer TAPS \cite{novotny:taps,gabler:response}.
A quasi-monochromatic photon beam was produced via bremsstrahlung tagging.
The photon energy covered the range 205--820 MeV with an average energy
resolution of 2 MeV.
The TAPS detector consisted of six blocks each with 62 hexagonally shaped
BaF$_2$ crystals arranged in an 8$\times$8 matrix and a forward wall
with 138 BaF$_2$ crystals arranged in a 11$\times$14 rectangle.
Each crystal is 250~mm long with an inner diameter of 59~mm.
The six blocks were
located in a horizontal plane around the target at angles of
$\pm$54$^{\circ}$, $\pm$103$^{\circ}$ and $\pm$153$^{\circ}$ with
respect to the beam axis. Their distance to the target was 55~cm and the
distance of the forward wall was 60~cm.
This setup covered $\approx$40\% of the full solid angle.
All BaF$_2$ modules were equipped with 5 mm thick plastic detectors for the
identification of charged particles.
The liquid hydrogen target was 10~cm long with a diameter of 3~cm.
Further details are described 
in \cite{novotny:phoswich}.

The measurement of the \Rpigp{} reaction channel was exclusive since
the 4-momenta of all
particles in the final state were determined.
The \Piz{} mesons were detected via their two photon decay channel and
identified in a standard invariant mass analysis from the measured photon
momenta.
The two \Piz{} decay photons and the \gprim{} photon in the final state
were distinguished by using
the \Piz{} invariant mass as a selection criterion.
The two photons with an invariant mass closest to the \Piz{} mass were
assigned to be the decay photons.
The protons were identified using the excellent time resolution of the TAPS
detector and the deposited proton energy:
The characteristic time of flight dependence on the energy of the
proton and a pulse shape analysis \cite{novotny:phoswich}
were sufficient to identify the
proton uniquely.
The proton energy calibration was performed by exploiting energy
balance of the exclusively measured \Rpip{} channel, thereby
compensating for
the energy loss in the target and plastic detectors.
Random TAPS - tagging spectrometer coincidences were
subtracted using background events outside the
prompt coincidence time window.

\begin{figure}
   \includegraphics[width=0.99\columnwidth]{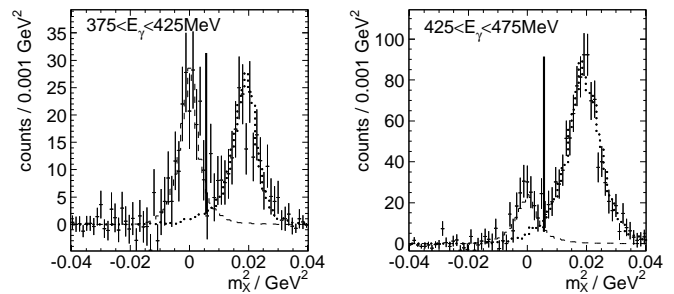}
\caption{Missing mass of the (\Piz $p$) system in the final state,
but with an additional photon detected for two different incident photon
energies.
The peak near 0.02 GeV$^2$ originates from 2\Piz{} production
and is cut away.
The peak at 0 GeV$^2$ shows the true \pigp{} production.
The dashed and dotted lines show the corresponding simulated
lineshapes using GEANT3.}
\label{fig:kin}
\end{figure}

Further kinematic checks were performed by exploiting the kinematic 
overdetermination of the reaction. Special attention had to be paid
to background from 2\Piz{} production 
arising from events in which one of the four 2\Piz{}
decay photons escaped detection due to the 
limited solid angle coverage of the detector.
In a first step, the conservation of the total momentum was checked in the
three cartesian directions respectively. After that, a missing mass
analysis was performed to discriminate the 2\Piz{} contamination.
The following missing mass was calculated:
\begin{eqnarray}
    M_{X}^2 = && ((E_{\Piz{}}+E_p)-(E_{beam}+m_{p}))^2 \nonumber \\
    && - ((\vec{p}_{\Piz{}}+\vec{p}_p)-(\vec{p}_{beam}))^2
    \label{eq:mmiss}
\end{eqnarray}
where $E_{\Piz{}}, \vec{p}_{\Piz{}}, E_p, \vec{p}_p$ denote the energy
and momenta of the \Piz{} and proton in the final state and $m_{p}$ the
proton mass. The resulting distributions (Fig.~\ref{fig:kin}) show two
distinct peaks, the widths of which are determined by the detector
resolution. The peak near 0.02~GeV$^2$ reflects the missing mass of a \Piz{}
and therefore originates from the 2\Piz{} production, while the peak
at 0~GeV$^2$ indicates the missing mass of a photon and hence the
\pigp{} production.
 A Monte Carlo simulation 
of the 2\Piz{} and \pigp{} reactions using GEANT3 \cite{geant} 
reproduces the lineshape of the measured data.
The nearly background free identification of the \Rpigp{} reaction is
demonstrated in Fig.~\ref{fig:kin}.
The remaining small
2\Piz{} background due to the finite detector resolution 
(16\% in the highest energy bin) 
is subtracted for the cross section determination.
Since the information of the photon \gprim{} has not been used for
evaluating the missing mass defined in Eq.~\ref{eq:mmiss}, another 
kinematic check has to prove that the photon \gprim{} is not
accidental. Therefore the energy balance was calculated to test
energy conservation:
$
    E_{BAL} = (E_{beam}+m_{p}) - (E_{\Piz{}}+E_p+E_{\gamma^\prime})
$
; the notation is the same as in Eq.~\ref{eq:mmiss}. The energy balance
confirms the clean identification of the \pigp{} reaction channel.

\begin{figure*}
  \includegraphics[width=1.5\columnwidth]{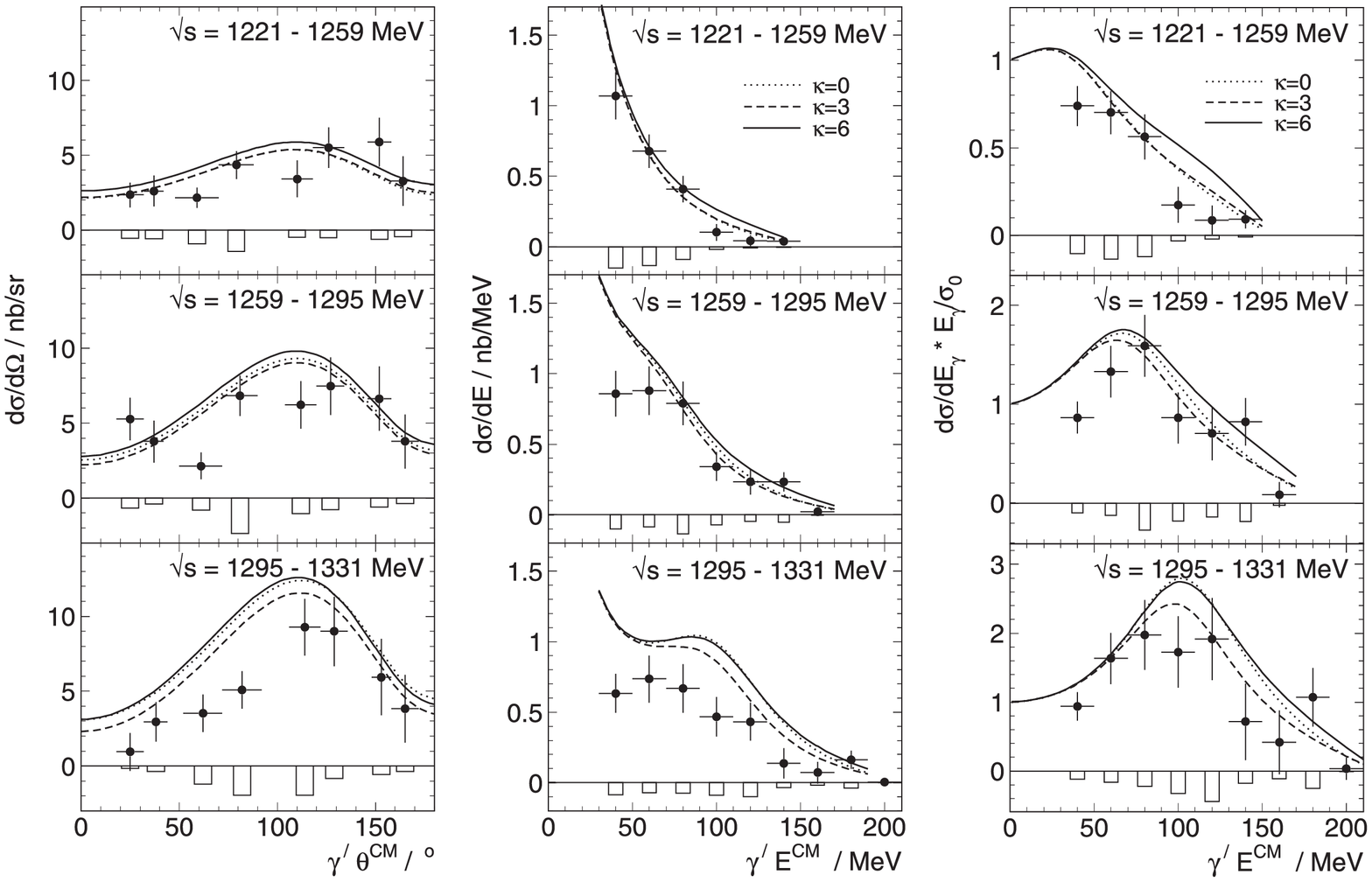}
  \caption{Differential cross sections for three different incident
excitation energies $\sqrt{s}$ in the CM frame. The systematic errors are
shown as a bar chart. \textit{Left:} angular
distribution of the photon $\gamma^\prime$; \textit{middle:} energy
distribution. 
The lines show the calculation
\cite{drechsel:pi0gp} for three different values of the 
anomalous magnetic moment $\kappa_{\Delta^+}$ = 0, 3 and 6.
On the \textit{right} side, the energy distribution has been
divided by the prediction of the soft photon limit
$\frac{\sigma_0}{E_\gamma}$, respectively for the data and the calculation.
 }
  \label{fig:cross}
\end{figure*}

The cross section was deduced from the rate of
the \pigp{} events divided by the number of hydrogen atoms
per cm$^2$, the photon beam flux, the branching ratio of \Piz{} decay into two
photons, and the detector and analysis efficiency.
The intensity of the photon beam was determined by counting the
scattered electrons in the tagger focal plane and measuring the
loss of photon intensity with a 100\%-efficient BGO detector 
which was moved into the
photon beam at lowered intensity. 
The geometrical detector acceptance and
analysis efficiency due to cuts and thresholds 
were obtained using the GEANT3 code and an event
generator producing distributions of the final state particles
according to \cite{drechsel:pi0gp}.
The systematic errors of the efficiency determination are small
because the shape of the measured distribution is reproduced by the simulation.
The average value for the detection 
efficiency is 0.25\%.

The measured differential cross sections for the
reaction \Rpigp{} are shown in Fig.~\ref{fig:cross} 
for three different incident excitation energies $\sqrt{s}$ (i.e. the
total $\gamma p$ center of mass energy),
starting at the $\Delta$ resonance position and going up
to 100 MeV above it. The angular distribution of the photon $\gamma^\prime$
in the CM system shows an enhancement for angles around
120$^\circ$. The energy distribution shifts towards 
higher $\gamma^\prime$ energies with rising $\sqrt{s}$, showing 
an 1/$E_\gamma$ form with an additional peak, where the strength and
the position depend on the 
excitation energy $\sqrt{s}$. The different reaction mechanisms suggest such
a behavior, where the 1/$E_\gamma$ dependence stems from the external
bremsstrahlung of the proton in the final state. The position of the 
peak structure (the energy of $\gamma^\prime$)
originating partly from the $\Delta$ radiation
is determined by
the difference of $\sqrt{s}$ and the $\Delta$ peak
mass and a small correction due to the available phasespace.
The $\Delta$ decay mechanism contribution is emphasized,
when the energy differential cross
section is divided by 1/$E_\gamma$ (compare the
column on the right hand side of Fig.~\ref{fig:cross}).

The first series of calculations, including only the resonant
$\Delta \rightarrow \Delta \gamma^\prime$ process 
as indicated in Fig.~\ref{fig:m1gamma}, were done by 
Machavariani et al.
\cite{macha:pi0gp, macha:pi0gp_err} 
and Drechsel et al. 
\cite{drechsel:pi0gp1}.
Both groups 
use the effective Lagrangian formalism and in addition the latter group uses  
a quark model approach to describe the reaction.
Since these calculations consider only the Feynman diagram which is
sensitive to \muD{}, they 
cannot reproduce the measured cross sections.

Recently, Drechsel and Vanderhaeghen \cite{drechsel:pi0gp} 
extended their model
and included bremsstrahlung diagrams (resonant $\Delta$,
non-resonant Born diagrams and $\omega$ exchange).
This calculation is shown in comparison to the measured cross sections in
Fig.~\ref{fig:cross}. The overall shape is reproduced very well,
although the absolute value is overestimated for the highest excitation
energy. This is related to a overestimate in the calculation
of the reaction \Rpip{}, which is well understood
and attributed to
$\pi N$ rescattering contributions \cite{drechsel:pi0gp}.
A model independent determination of the \pigp{} cross section is
feasible in 
the soft photon limit, which relates \pigp{} production to
\Piz $p$ production in the limit of vanishing photon energy $E_{\gprim}$ 
\cite{marc:lowenergy}:

\begin{equation}
 \lim_{E_{\gprim}\rightarrow 0} 
 \biggl({\frac{d\sigma}{dE_{\gprim}}}\biggr)  =  
 \frac{1}{E_{\gprim}}\cdot {\sigma}_{0}
 \label{eq:soft_photon} 
\end{equation}
\begin{align}
 {\sigma}_{0} = &
 \int{\!d\Omega_{\Piz} \biggl({\frac{d\sigma}{d\Omega_{\Piz}}}\biggr) }
 \! \cdot \! \frac{2\alpha_{em}}{\pi}
 \left\{%
 {\biggl(\frac{v^2 \! + \! 1}{2v}\biggr)
 ln \biggl( \frac{v \! + \! 1}{v \! - \! 1}} \! - \! 1 \biggr) 
 \right\} 
  \label{eq:low_energy} \notag \\
 v = & \sqrt{1-\frac{4m_p^2}{t} } \quad , \quad  
 t = (k - p_{\Piz{}})^2 \notag
\end{align}
$d\sigma / d\Omega_{\Piz{}}$ labels the differential cross section for
\Piz $p$ production, $m_p$ the proton mass, $t$ 
the four momentum transfer between the initial photon and the \Piz{}
meson and $\alpha_{em} = e^2 / 4\pi \approx 1/137$.
According to Eqn.~\ref{eq:soft_photon},
the energy differential cross section divided by
$\sigma_{0}$/$E_{\gprim{}}$ should be equal to 1 in the limit of
zero photon energy $E_{\gprim}$. This ratio is shown in the right
column of Fig.~\ref{fig:cross}, where the differential
cross section $d\sigma / d\Omega_{\Piz{}}$ 
in Eqn.~\ref{eq:soft_photon} is calculated with the same
effective Lagrangian model \cite{drechsel:pi0gp}. For comparison to the
experimental results, the data are also plotted as a cross section
ratio where $\sigma_{0}$ has been determined from
Eqn.~\ref{eq:soft_photon} using consistently the 
measured differential
cross section $d\sigma / d\Omega_{\Piz{}}$ of the \Rpip{} reaction.
The cross section ratios show better agreement; they are less
sensitive to uncertainties in the model calculation as well as
uncertainties in the determination of the photon flux and target
length.

The sensitivity to the magnetic moment of the
$\Delta^+$ is illustrated in Fig.~\ref{fig:cross}
 by the difference of the three curves. The $\Delta^+$
magnetic moment can be obtained from the anomalous
magnetic moment $\kappa_{\Delta^+}$ which is the only free parameter 
of the calculation \cite{drechsel:pi0gp}
\begin{equation}
  \mu_{\Delta^+} = (1+\kappa_{\Delta^+}) \cdot \frac{e}{2m_\Delta} =
    (1+\kappa_{\Delta^+}) \cdot \frac{m_N}{m_\Delta}
      \cdot \mu_N 
  \label{eq:kappa_mu}
\end{equation}
where $\mu_N = e / 2m_N$ is the nuclear magneton.
A combined maximum likelihood analysis \cite{pdg_a:2000} of the three 
cross section ratios in Fig.~\ref{fig:cross} yields a value of
\muD{} = $(2.7_{-1.3}^{+1.0} \pm 1.5)\mu_N$,
the goodness of fit is 
$\chi^2/F = 1.8$ (F=21).
The first error represents the statistical uncertainty and the
second one reflects the systematic errors given in Fig.~\ref{fig:cross}.
This error does not include the systematic error of the model calculation
which is of the order of  $\pm 3 \mu_N$, estimated from the uncertainties
discusssed in \cite{drechsel:pi0gp}.
The extracted value of \muD{} is in the range
of different baryon structure
calculations \cite{leinweber:muD,butler:muD,kim:muD,aliev:muD},
but not sensitive enough to discriminate between them.

In conclusion, we have made the first measurement of 
the magnetic moment of the
$\Delta^+(1232)$ resonance by exploiting the
reaction \Rpigp{}.
We see a clear deviation from a soft bremsstrahlung cross section
at higher energies of the radiated photon, pointing
to a sentitivity to the magnetic moment of the $\Delta^+(1232)$ resonance.
However,
the limited statistics and the 
uncertainty of the model lead to a value of the $\Delta^+(1232)$
magnetic
moment, which is not sufficiently precise
for a detailed test of different baryon structure calculations.
This situation calls for a
follow up experiment with much higher statistical precision,
using 
so that the 
kinematic regions most sensitive to \muD{} can be exploited
\cite{drechsel:pi0gp, kottu:mdm_taps}. 
An investigation of the cross section asymmetry using a polarized 
photon beam would also be valuable. 
Supplementary, an improvement in the theoretical description
is necessary to minimize the model dependence.
The measurement can be extended to higher excited
states of the nucleon. In particular the magnetic
moment of the
S$_{11}$(1535) resonance is accessible via the reaction
$\gamma p \rightarrow \eta \gamma^\prime p$ because of its clean
distinction from other resonances in the second resonance region through the
$\eta$ channel.

It is a pleasure to acknowledge inspiring discussions with M. Vanderhaeghen,
D. Drechsel, A. Machavariani and A. Faessler. 
We would like to thank the accelerator group of MAMI 
as well as many other scientists and technicians of the Institut fuer
Kernphysik at the University of Mainz for the outstanding support.
This work is supported by DFG Schwerpunktprogramm:
"Untersuchung der hadronischen Struktur von Nukleonen und Kernen mit
elektromagnetischen Sonden", SFB221, SFB443, the UK Engineering and Physical Sciences Research Council and Schweizerischer Nationalfond.

\bibliography{./kotulla_mdm.bbl}

\end{document}